\documentclass[aps,a4paper]{revtex4}

\begin{document}

\title{Quantum Biology}
\author{Alessandro Sergi}
\email{sergi@ukzn.ac.za}

\affiliation{School of Physics, University of KwaZulu-Natal, Pietermaritzburg,
Private Bag X01 Scottsville, 3209 Pietermaritzburg, South Africa}

\begin{abstract} 
A critical assessment of the recent developments
of molecular biology is presented.
The thesis that they do not lead to a conceptual
understanding of life and biological systems is defended.
Maturana and Varela's concept of autopoiesis is briefly sketched
and its logical circularity avoided by postulating
the existence of underlying {\it living processes},
entailing amplification from the microscopic to the macroscopic scale,
with increasing complexity in the passage from one scale to the other.
Following such a line of thought, the currently accepted model of 
condensed matter, which is based on electrostatics and short-ranged forces,
is criticized. It is suggested that the correct interpretation
of quantum dispersion forces (van der Waals, hydrogen bonding, and so on)
as quantum coherence effects hints at the necessity of including
long-ranged forces (or mechanisms for them) in
condensed matter theories of biological processes.
Some quantum effects in biology are reviewed
and quantum mechanics is acknowledged 
as conceptually important to biology since without
it most (if not all) of the biological structures
and signalling processes would not even exist. 
Moreover, it is suggested that long-range
quantum coherent dynamics, including electron polarization,
may be invoked to explain signal amplification
process in biological systems in general.

\vspace{0.1cm}

\noindent
{\bf Published on line in:}\\
{\it Atti della Accademia Peloritana dei Pericolanti}
Vol. LXXXVII, C1C0901001 (2009).\\
DOI: 10.1478/C1C0901001
\end{abstract}

\maketitle

\section{Introduction}  
Biology offers to scientists the most complex systems to study in the universe. 
Since scientists themselves are biological systems, such a study
is not just the most interesting that one can think of
but effectively introduces a circularity in the process of knowledge,
as was noted by Maturana and Varela~\cite{tree}: Life systems (the scientists)
who try to know life systems (possibly themselves); in other words,
life that tries to know life.
It is a platitude to assert that studying biology from the point of view of
physics (i.e., the point of view of the fundamental laws of the universe)
is very difficult~\cite{st}. From a physicist's perspective, there are universal laws
(and, perhaps, building blocks) of reality and one should be able to predict
the emergence and the characteristics of biological systems from these very
fundamental laws. Such a gigantic endeavour has not been successful to date.

In this contribution, the thesis that quantum mechanics is a powerful tool
for explaining the characteristics of biological systems will be defended
and some (speculative, at the moment) lines of research will be suggested.
From a certain point of view, the use of quantum mechanics in biology
might seem logical since it is the fundamental theory describing
microscopic phenomena in physics. 
Within a fully reductionist philosophy (which it is not 
invoked here) chemistry, biochemistry, and biology would be only epiphenomena
of the fundamental microscopic laws of physics, i.e., they would be secondary 
manifestations of the main microscopic reality with its laws.
From another point of view, it is very strange that one would invoke 
such a controversial theory, as quantum mechanics actually is, in order to explain
the most complex phenomena in the universe. As a matter of fact, while
everybody agrees on the {\it main} technical points of quantum mechanics,
almost nobody agrees on its interpretation, which seems to depend oddly
on the area of research the theory is applied to.
It is not entirely wrong to write that there are almost as many interpretations
of quantum mechanics as there are theoretical physicists: It is sufficient
to search in the contemporary literature of physics journals to be convinced
of this.
The ideas underlying this contribution are that quantum mechanics 
is the fundamental physical theory in the microscopic world,
that important concepts can arise from its application 
to biological systems, and that biological systems, while requiring
a multi-level approach, must also be studied 
from a microscopic point of view (in order to unfold 
their universal characteristics).

The main-stream scientific discipline that currently undertakes
the endeavour of a microscopic explanation of life is molecular biology. 
Its method is based on enumerating and elucidating the role
of molecules in the life process. While such a work is
of fundamental importance for medicine and applied biochemistry,
it does not seem to lead to a better understanding
of the universal properties of life itself.
Such a critic opinion, which has been defended
among others by Kaneko~\cite{kaneko}, will be the starting point of this contribution.
However, while Kaneko invokes dynamical system theory~\cite{kaneko},
here quantum mechanics is considered necessary
in order to unveil the universal properties of living systems.

This contribution is organized as follows.
In Section~\ref{sec:whatislife} the arguments of Kaneko~\cite{kaneko},
trying to characterize life and to criticize molecular biology
(for failing to provide an understanding), are followed.
In Section~{\ref{sec:autopoiesis}} the universal logic of Maturana
and Varela, which is founded on the concept of autopoiesis,
is summarized. Autopoiesis is further analyzed
and the more fundamental concept of {\it living process}
(based on amplification mechanisms) is introduced. 
We feel that \emph{living processes}
can be directly linked to quantum phenomena
and, as such, are more suited to physical modeling. 
The current paradigm of condensed matter physics
is illustrated and criticized in Section~{\ref{sec:cmp}}.
The features of quantum mechanics, with a particular emphasis
on those of interest to biological processes,
are sketched in the same Section.
In Section~{\ref{sec:bioqm}} a certain number of biological
phenomena, where quantum mechanics is necessary, are reviewed.
Van der Waals interactions 
and quantum mechanical dispersion forces are presented 
(in a speculative way) as the main candidates
for the amplification processes necessary to living systems.
Finally, conclusions and perspectives are elucidated 
in Section~{\ref{sec:conclusions}}.


\section{What is life?}\label{sec:whatislife}

The questions that will be addressed in this Section are:
``What kind of system is life?'' and
``What does {\it understanding life} really means?''.
The main thesis is that
such basic questions on life systems 
are not answered by the main-stream
approach of current biology, which enumerates molecules and genes.

As for understanding life as a process, the molecular paradigm 
embraced by contemporary biology
has a fundamental flaw:
There is no particular molecule, including DNA,
whose presence by itself implies life.
In his book~\cite{kaneko}, Kaneko presents the semi-serious example
of an omelette, which possesses DNA but which is clearly not alive.
Moreover, until one does not gain a general understanding of what life is,
speculations about the possibility
that the molecules, used by living creatures
on Earth, could not be the only thing 
playing an important role for life are not entirely unreasonable.
Hence, if not the molecules, the specific conditions for life
remain to be clarified.

One may attempt to compile a list of the
characteristics of living systems. These are the ability of reproduction,
the potentiality to undergo evolution,
the existence of some kind of structure
separating an individual's body from the external world,
some kind of metabolic capacity through which
an individual body is maintained,
the existence of some degree of autonomy, and so on.
Despite the various attempts at listing such characteristics,
no list has ever been completely satisfactory or agreed upon.
However, there must be a solution since in many cases human beings
have an intuitive ability to distinguish between living 
and non-living creatures (excluding limit cases such as viruses and so on).
Acknowledging that certain properties are common to all living systems,
a theoretical physicist would like to 
search for the universal properties of living systems.
In other words, one would like to understand the
universal logic of life (its {\it logos} underlying it
as a process) instead of understanding the specific functioning
of a definite organism.
Indeed, since its birth biology has attempted to escape
mere enumerationism through Darwin's theory of evolution:
A universal logic based on the three processes 
of variation, reproduction, and selection.
Such an evolutionary logic has been mimicked by computer scientists
in order to devise the so-called {\it genetic} algorithms.
However, Darwin's theory alone does not allow one to determine what kind
of properties (or functions) of organisms are possible in general, nor does it
allow one to determine whether any specific property 
can be realized in practice.
Hence, a universally applicable logic explaining
the emergence of the fundamental properties
of living systems is yet to be found.

\subsection{Molecular biology}

Physicists introduced a major trend in biology more than half a century ago.
It is worth mentioning Delbruck and collaborators,
who were strongly influenced by the lectures of Niels Bohr.
Perhaps, the most far-reaching speculation is due
to the father of (quantum) wave mechanics,
Schr\"odinger himself, who suggested that
an aperiodic solid could be the ``storing device'' 
for biological information,
in his famous book ``What is life?''~\cite{whatislife}. 
This steered the search leading to the discovery of the DNA molecule
by Watson and Crick.

Since the discovery of DNA,
molecular biology has attempted to describe the universal
properties of the phenomena exhibited by living systems
in terms of molecules.
The goal was, and still is, to trace down chemical processes
from the level of cells
to that of the composing molecules, and to understand
the functioning of each molecule in biological processes
(e.g., heredity, metabolism, motility, and so on). 
The methodology of molecular biology can be sketched in the following way.
First, one has to consider a system at the macroscopic level and
identify the molecules and genes that are important in some function 
under study. The role of each molecule must be clarified 
and the interactions of such molecules with other molecules must be found.
Then, one has to devise how the macroscopic functions of the organism
arise from the cooperativity of the microscopic relevant molecules.
In order to bring such a program to completion, one has to
cope with the enormous combinatorial complexity that is due
to the great number of different types of molecules involved
and, nevertheless, devise the network/circuit of chemical reactions 
and back-reactions entailing life as a process.
Hence, molecular biology originally started as the pursuit of universality,
rejecting the ``enumerationism'' that preceded it.
However, the present days witness a
re-emergence of the enumerative doctrine of the past,
even if in different and more subtle forms.
Indeed, under the push of gigantic funding from medical
(and perhaps, army) research companies, molecular biology
has now become an enumerative science again.
One can classify the
genome project (the listing of all the human genes),
the proteome project (the listing of all proteins),
and the metabolome project (the listing of all the molecules
involved in metabolism) as mere enumerative science.
Of course, such projects are of utmost importance for practical reasons
such as the health care of human beings.
The point is that such projects alone do not lead scientists
one inch further in the understanding of the universal logic of life.

Essentially, reductionism is 
the philosophy underlying such enumerative projects. 
However, there are some hidden assumptions behind reductionism
that need to be brought to the foreground.
Typically, reductionism is based on the premise that the properties
of individual elements change little in response
to their interaction with the other elements composing the whole.
Here one faces a first problem because
interactions are often not small in biological systems.
Kaneko illustrates the example of
proteins in the crowded environment of a living cell,
where they effectively constitute a gel~\cite{kaneko}.
In some cases, the distance between two neighbouring atoms in a
single protein molecule is greater than that between either of these
atoms and the closest atoms of other protein molecules.
In reality, it is not unreasonable to believe
that a hard-core version of reductionism is bound to fail
in the search for the universal logic of life
simply because living systems are not machines.
Typically, in living systems the behaviour of
the parts/molecules alone is different from that of the parts/molecules
acting collectively.
In other words, the dynamics of
the parts composing the whole is determined by the whole;
an example is given by the process of morphogenesis.
In addition, living systems display no fixed response 
to a specific stimulus:
A given stimulus  can be associated to various possible responses
(a mathematical analogy is provided by many-valued functions).
Technically, such a variety of responses to a fixed stimulus
is referred to as absence of {\it stiffness}. 
Such an absence of stiffness can be further analyzed
in terms of {\it softness}, or the dependence
of the response on the environmental conditions,
and of {\it spontaneity}, or the possibility
of associating different outputs to the same input,
depending on the internal state of the living system
and its fluctuations.
Softness and spontaneity, together with some form of memory,
give rise to (perhaps) the most striking feature of living systems:
Autonomy (flexibility and adaptability).
In other words, living systems do not always behave 
in strict accordance with the rules applied to them
and, depending on the situation, the rules they perceive will change
(or the living systems are able to change the rules they abide).
Linking the autonomy of living systems to mere molecular processes
seems impossible to the present author.
Other universal properties of living systems that cannot 
be traced back to molecules alone~\cite{kaneko}
are the stability, the irreversibility in the development process
(i.e., the loss of multipotency from embryonic stem cells
to stem cells, and to committed cells capable of
reproducing their own kind only), and
the compatibility of the two faces of the reproduction:
The ability to produce nearly identical offspring
and the capacity to generate variations leading to diversity
through evolution.

\subsection{The information paradigm}

Since this is the computer age, it is not strange
to witness another more subtle reductionist approach
to the logic of life by means of the
the information (computer) paradigm.
Such an attempt is subtle because, in principle, it promises
to explain even the autonomy of living systems.
Typically, within an algorithmic approach,
the process of development exhibited by living creatures
would be represented as some kind of computer program
(a logical expression in the form of a chain of if-then
statements). For example, if $X$ denotes the concentration
of a species of molecules and $Y$ the concentration of another species,
a fundamental bio-chemical process
could be understood by the  code:
$${\rm If}~ X ~\rangle ~X^0 ~{\rm then~ express}~ Y\;,$$
where $X^0$ is a concentration threshold. 
One serious problem that such an algorithmic approach must face is
that, for continuous values of the concentration $X$,
the fluctuations of $X$ can cause big errors and
one should then explain the 
algorithmic robustness (stability) of such ``living'' computers.
The problem of stability is not trivial even when one
considers
discrete values of $X$.
In fact, life processes involve an enormous variety of molecules.
But, the number of molecules of a given type is often small.
This means that the fluctuations of $X$ are always big
and one is in the presence of a very large chemical noise. 
An example is given by
the process of development of a multicellular organism~\cite{kaneko}:
It seems miraculous that systems with such a variety of molecules,
participating in such a large number of processes,
can result again and again in an almost identical macroscopic pattern.
The situation is analogous to that of a person attempting to stack many 
irregularly shaped blocks into a particular form during
an intense earthquake~\cite{kaneko}.
A possible solution to the stability problem might be given
by the existence of  negative feedback processes in living systems
(i.e., by some form of error-correction).
However, there must also be positive feedback (amplification) in
living systems. Indeed, in the following, the thesis that
the amplification processes themselves are a key to the understanding
of life will be proposed.
It is not clear to the author how error-correction schemes 
might work in the presence of amplification.
However, for the sake of scientific fairness, one should acknowledge
that the above arguments alone are not enough to
rule out completely the information paradigm.
Perhaps, it is the author's dislike of such a paradigm 
that leads him to reject it.
Typically, if the universe is some kind of supercomputer
and the phenomena we observe are just the results of its calculations,
then everything should be computable by the computer-universe.
However, according to the laws we know, 
many non-linear problems show an extreme sensitivity
to the initial conditions, and thus require 
an infinite precision representation of the initial information for an
exact solution. There are even phenomena which are not computable 
in a finite time with a finite memory.
In classical mechanics the three-body problem is not exactly solvable,
in quantum mechanics the two-body problem (e.g., electron dynamics
in the helium atom) is not exactly solvable,
and in quantum field theory the zero-body problem (i.e., the vacuum) is not 
exactly solvable. Scientists are able to calculate realistic processes
only at the expense of tremendous simplifications and approximations.
The author is not able to imagine what kind of computer-universe
would be able to solve all these problems in the time they take
to occur in reality.


\subsection{Autopoiesis: The universal logic 
of Maturana and Varela}\label{sec:autopoiesis}

Molecular biology is not the only attempt to provide
a universal logic for life. In recent times, Maturana
and Varela proposed a theory of life that
is not based on molecules
and that tries to explain the general features of living systems,
including those features that are not observed but which are nevertheless
possible, in abstract terms~\cite{tree}.
These authors state that
understanding a living system means understanding the
network of relations that must occur so that it can exist
as a unit. The set of such relations is called ``organization''.
The particular and concrete realization 
of the organization
of a living system (molecules, network of specific chemical reactions,
and so on) is called ``structure''.
From this distinction, it follows that
the same organization can, in principle,
be realized through various structures.

These authors make another very important distinction.
In their analysis~\cite{tree} it emerges that
living systems are closed (e.g., isolated) from the point of view
of their organization. In other words, whatever the logic of life may be,
it must specify the living system as an independent unit,
knowing and reacting to nothing else than its internal state.
However, Maturana and Varela also clarified that, on a different level,
viz., the level of thermodynamics, living systems are 
open systems: They are not isolated
but (from the point of view of their structure)
interact continuously with their own environment.
At this stage, Maturana and Varela explain
the process of life in terms of an unavoidable circularity:
It is a peculiar feature of living systems that the product
of their organization is themselves.
In other words, their organization is such that it maintains
their structure which, in turn, implements in practice their
organization.
Such an unavoidable circularity is called 
autopoiesis, and Maturana and Varela proceeds with the identification
of  living systems with autopoietic units.

Clearly, the logic of autopoiesis is a universal logic,
based on the circularity of the abstract level
of the organization maintaining itself by means of
the concrete (and physical) level of the structure
of living systems. A platonic philosopher
might, perhaps, use the word ``form'' in place of organization,
hence clarifying from the very begining that Maturana and Varela's
approach is not entirely reductionist.
Nevertheless, reductionism is to be found at the level
of the structure, the concrete physical realization
of the living system which must be invoked in order to maintain
the organization itself.
Such a circularity is fascinating and problematic at the same time
since, in general, physics and human logic do not like circularity.
It seems difficult to devise concrete mathematical models
implementing such general ideas.

\subsection{Living processes}\label{sec:livproc}

One way to escape the circularity of autopoiesis would be
to decompose it in terms of more fundamental processes.
Indeed, it appears to the present author that
autopoiesis is necessary to explain the persistence in time
of a given living system: The system can stay alive because its actions
mantain its existence.
The following question naturally arises: Would it be possible
to speak of transient living systems?
In other words, if living systems would not produce themselves
they would disappear almost instantaneously; however, 
it is altogether tempting to call ``life'' such an ephemeral existence.
I propose to consider {\it living processes}
(although the meaning of this concept is at the moment unclear)
as the fundamental building block of life.
Such living processes would be something more fundamental than autopoiesis
for characterizing life.
From such a point of view,
autopoiesis would explain the stable existence of a living
system over an extended time interval.
it is intuitive that when autopoiesis stops, a living system dies.
Hence, an autopoietic unit would then be defined as a network 
of self-sustaining living processes. 

Although the living process has been defined as the fundamental 
(transient) building block of life, its characteristics have been left 
as yet
mysterious. The working hypothesis that is introduced here 
as a postulate
(to be verified by further analysis) is that
the living process is an amplification process,
from the microscopic to the macroscopic scale, which builds up
complexity  in structure and organization. 
Of course, the autopoiesis of Maturana and Varela would require
a feedback process from the macro to the micro scale.
However, according to the above discussion, such a feedback mechanism
would be required for mantaining life, not for life itself. 
The introduction of the concept of living process breaks the circularity
of autopoiesis. The identification of the living process
with a particular type of amplification process, transferring information
from the microscopic level to the macroscopic one, which also leads
to an increase of complexity in organization, permits, at least in principle,
to devise structural (i.e., physical or mathematical) models.
At this stage, it is also necessary to explain what we mean when saying 
that complexity increases in going from smaller to larger scales.
Complexity is the number of constraints (or laws)
that the process must fulfill as the scale increases.
The nature of such constraints can be static or dynamic, and one
thus considers a fixed structure or to the time evolution of the system.
It is not difficult to understand that an increasing number
of constraints leads to the generation of forms: What is form
if not something that is specified by boundaries and constraints?
Therefore, one can also define the living process as an amplification
process that builds up forms.

The above reflection is, at the moment of writing, only a working 
hypothesis and here the author
is not going to provide the reader with any specific mathematical
or computer model of a living process.
Perhaps, the best that can be presently done is
leaving the reader with a metaphoric image that tries to convey
the idea of amplification building complexity.
Hence, one can imagine a lightning flash which, from the shape of 
a flux-tube, enlarges (amplification), not unlike the delta of a river,
in order to build up an intricate tree (augmented complexity) of
smaller lightning flashes.
Keeping the poetic spirit of the above example,
one can draw the main conclusion
of this Section and state that biological systems, 
far from being mere machines,
are {\it matter that dances} (i.e., matter that
moves with an incredible level of coordination among its constituents).
Theoretical physicists are left with the question whether
the standard theories of condensed matter physics 
can explain biological systems.


\section{The current paradigm of condensed matter physics}\label{sec:cmp}
The physical description of condensed matter systems
is currently dominated by electrostatics.
The most sophisticated, and state-of-the-art, molecular dynamics simulations
of protein molecules in water, see Refs.~\cite{sergi-rome,pellicane} as an
example, are based on semi-phe\-no\-me\-no\-lo\-gi\-cal force fields, 
describing interactions arising from fixed electric charges, Lennard-Jones
and harmonic potentials plus bond constraints that mimic covalent bonding.
In practice, charge shielding causes the existence of short-ranged forces
in such models, which effectively treat matter as an {\it erector set} 
(or meccano).
Even first-principles theories, such as the electron Density Functional Theory
~\cite{gross}, are currently based on electrostatics alone.
As a result, they describe hydrogen bonding, van der Waals, and 
charge polarization effects only with difficulty and, on the whole, with 
unsatisfactory results.
Clearly, such a condensed matter paradigm tries
to build long-ranged correlations from statistical fluctuations of 
short-ranged interactions.

We surmise that such a paradigm might be flawed at a fundamental level.
For example, it is clear that both the structure and the function of 
biological macromolecules
largely depend on hydrogen bonding as well as on hydrophobic and 
hydrophilic interactions.
These are determined by dispersion or van der Waals forces, 
precisely those interactions that are not properly described by the current
paradigm.
Such dispersion forces depend on the temperature 
and on the molecular environment
(i.e., they are not additive)~\cite{milonni,vanderwaals}.
They lead to the existence of long-ranged networks of structural and 
dynamical correlations.
Van der Waals forces, also known as induced-dipole-induced-dipole forces,
arise from a highly correlated motion of the electronic clouds
of otherwise neutral atoms. Such a correlated motion, 
which in quantum mechanical terms is called {\it coherent},
takes place even at room temperature and in densely packed matter.
Poetically, one could say that such forces in matter arise from
``dancing'' electronic clouds.
The explanation of such a dance is provided by
quantum mechanics~\cite{milonni,vanderwaals,ballentine}.


Quantum mechanics is widely believed to be the fundamental 
theory underling the phenomenological reality.
Although the majority of physicists agrees on its mathematical
formulations, its interpretation is highly controversial.
However, on some points there is a wide consensus.
For example, there is almost no dispute on the issue that
quantum mechanics has some form of non-locality built 
inside~\cite{bell}.
The most advanced mathematical formulation of quantum mechanics
takes the form of a field theory~\cite{zee}.
Notwithstanding infinities, field phenomenology
is perhaps more soundly funded than conventionalism 
or the {\it spooky} attitude arising from ``particle'' interpretations
(see, for example, the discussion of the Einstein-Podolski-Rosen
paradox from the point of view of field theory 
in Ref.~\cite{preparata}). In simple terms, the field is ``something''
that is extended in space and time by its very definition and,
from a conceptual point of view, this aspect can be accorded more easily
with the non-locality of 
quantum mechanics, which appears rather puzzling when interpreted 
in terms of localized particles.
What is important, both for physics and for the present discussion,
is that the quantum phenomenology (therein including
discreteness, diffraction, and coherence~\cite{ballentine})
does not substantially raise any dispute.

Discreteness in quantum mechanics is usually associated with
the appearance of stationary energy levels
separated by ``quanta'' of energy: Transitions between
these levels can only take place through the transfer of the required
amount of energy.
Such a discreteness arises from the boundary conditions
imposed on the wave function (or functional) and is considered
to be well understood.
The existence of discrete values for the magnetic moments,
charges, and masses of fundamental particles is less clearly understood
but is easily embedded in the current formalism of quantum mechanics.


Quantum diffraction takes its name from an analogy
with the wave propagation of light.
Ensembles of microscopic particles exhibit
wave-like motion arising from the correlation
and the spatial and time memory of single events
in an ensemble: One single particle
is able to influence the ``whole'' so that the entire ensemble appears
to ``move'' like a wave, thus also displaying interference effects.


Coherence also takes its name from an analogy with the wave
motion of light: It immediately brings to mind the condition
of phase stability that is necessary to observe interference
and, thus, diffraction.
In quantum mechanics one would consider the phase stability of
the wave function (or functional in field theory). However, this
would be confined within a formulation of quantum mechanics
in terms of wave functions. Quantum mechanics can also be formulated
in terms of path integrals, charge and mass densities, distributions
in phase space and so on~\cite{nine}.
Hence, one needs a definition of quantum coherence less bound to the
mathematical formulation of quantum mechanics itself.
Here it is proposed to define quantum coherence
as the property underlying the typical and highly correlated motion which
takes place
in an unperturbed, isolated quantum system.
In condensed matter physics, striking examples of quantum coherent motion
are provided by superfluids and superconductive materials~\cite{zee}.
In superfluids the atomic motion is so highly correlated that friction
disappears and the fluid moves as a whole without dissipation.
In many respects, a superconductive material can be considered as
a charged superfluid of paired electrons (Cooper's pairs)
moving in a frictionless way in the lattice of positively charged
ions (making up the solid material).
Both phenomena take place at very low temperatures. Thermal fluctuations
are incoherent by their very essence and destroy coherent
quantum fluctuations with a surprisingly high efficiency:
This is the phenomenon of decoherence~\cite{decoherence}
that is displayed by open quantum systems~\cite{petruccione}.
Hence, in real systems some kind of shielding from thermal fluctuations
is necessary in order to observe quantum coherent motions.
It is worth remembering that in both cases of superfluidity 
and superconductivity, physicists do not possess a widely agreed
microscopic dynamical explanation.
In superfluids, a microscopic picture of rotons, which are the typical
many-body excitations of such systems, has not yet emerged.
In other words, although one can approximately calculate
the roton energy spectrum, it is not yet known what a roton is
on the microscopic scale: In other words, nobody actually knows what the 
{\it rotonic} motion of atoms in a superfluid is.
The situation is somewhat better for superconductivity,
where, at least, Cooper's pairs have been postulated.
However, there is no first-principle explanation
of the formation of Cooper's pairs. 
Typically, the celebrated Bardeen-Cooper-Schrieffer
theory of superconductivity~\cite{ballentine,zee} works only {\it after}
assuming the phenomenon of electron pairing.
In analogy with mechanics, one can call such theories 
{\it kinematical}, since they describe the time evolution of the system
without considering the microscopic causes of the motion.
This state of affairs should be contrasted with that of {\it dynamical}
theories which try to explain the time evolution of the 
system under study starting from the underlying microscopic causes 
(in analogy with dynamics, the branch
of mechanics which explains motion in relation to forces).


As for its applications in 
condensed matter systems, quantum mechanics might provide 
a synthesis of reductionism and holism.
The big fight of the $19^{\rm th}$ century,
between Mach and Ostwald on one side (the champions of
thermodynamics and of the holistic vision of matter)
and Maxwell and Boltzmann (the champions of atomism) on the other,
has been resolved in favor of the latter:
The regularity of the chemical laws
is nowadays commonly seen as the realization
of the ancient atomistic dream of Democritus and Epicurus.
However, the importance of the specification of the boundary 
conditions in quantum mechanics, arising from its non-local
or global (holistic) features, renders it conceptually
similar to thermodynamics, where the ensemble must be specified in terms
of the macroscopic conserved quantities.

The main conclusion of this Section is that condensed matter systems
can be dynamically understood only by resorting to quantum mechanics
and to the coherent motion of the electrons, which gives rise to
chemical bonding:
Quantum mechanical electronic clouds are matter that dances,
the poet would say.
The main question that is left to theoretical physicists is the following:
Condensed matter requires the coherent motion of electrons while
biological systems seem to require a highly correlated motion
of heavy atoms;
is quantum mechanics necessary to
understand biological systems as dancing matter?
In other words, may coherent electronic motion be the cause
of correlated heavy atomic motion?

\section{Quantum phenomena in biology}\label{sec:bioqm}

From a certain point of view, it should not be a surprise that
quantum mechanics is relevant to biology.
After all, quantum mechanics is definitely relevant to chemistry.
However, there is the serious possibility that all quantum
effects in biology are, in practice, trivial~\cite{nontrivial}.
By this it is meant that, although necessary to explain the details
of a given biological phenomenon, quantum mechanics does not need to 
be understood mathematically by a biologist who wants to study life processes.
Up to a certain extent, such a position is also held in the present
contribution and phantasmal concepts such as 
``quantum consciousness''~\cite{penrose} are not even discussed.
Because of decoherence~\cite{decoherence}, it appears
highly implausible to the present author that extended coherent states
of heavy atoms, such as those
of superfluids and superconductors, may exist at room temperature
inside a biological system (such as a cell).
Is this the end of the story?
Since electronic coherence seems to be fundamental
for condensed matter systems (in practice all electromagnetic forces
can be seen as a manifestation of quantum coherent 
behaviour~\cite{collective}), there is still the possibility
that such an electronic coherence is really fundamental
biological systems.
Such a thesis is here defended and a working hypothesis is also proposed.

There are some quantum phenomena in biology
that definitely need quantum mechanics for their very existence
and whose detailed description is challenging 
for the theoretical physicist. However, once their existence
is postulated, they can be used by the biologist
without almost any reference to quantum mechanics itself
(in the jargon used in this contribution, one can say
that such phenomena are {\it kinematically} described
by the biologist). This occurrence is not very dissimilar
from the understanding of the stability of matter: Without quantum
mechanical laws atoms could not exist. However,
in a kinematical way, one can postulate the stability of atoms,
disregarding its cause, and study, for example, the physics
of noble gases (at temperatures far from the absolute zero).
On a second thought, perhaps it is not intellectually fair
to classify effects like these as trivial.
Typically, the fact that some fundamental concept can be used as a 
``black box'', within a more approximate level of description,
should not be used to deem the concept itself as trivial.
Therefore, in disagreement with the definition of triviality
adopted in Ref.~\cite{nontrivial}, here it is defended the thesis
that when quantum mechanics is necessary for the existence
of a given biological phenomenon, that phenomenon
is nontrivial on a quantum mechanical basis, even if biologists may choose
to describe it kinematically as a black box.
Electric charge and exciton transfer processes in proteins and 
photoactive complexes are examples
of such pseudo-trivial quantum effects.
The exciton is a many-body excitation of an interacting system:
A charge is excited from its ground state and undergoes a transition 
to the excited
state leaving a hole (a missing charge) in the ground state;
afterwards, because of many-body interactions, the charge and the hole
move in a coherent way thus giving life to the exciton.
Without quantum mechanics, one would not have ground and excited
energy states and would not have the coherent motion of the 
charge and the hole.
In biological systems, there are cases 
in which the motion of the exciton can be approximated
by classical mechanics (and there are cases when this is not possible)
but without quantum mechanics the exciton itself would not exist.
Single charge transfers are intrinsically quantum mechanical
only when tunneling through energy barriers takes place.
As a matter of fact, some charge transfer processes can be modeled
classically. However, quantum coherence could be fundamental also
in non-tunneling transfers in order to establish
the right degree of correlation with the environment
rearrangement before, during, and after the charge transfer.
An example could be provided by
the process of molecular recognition taking place
in odor sensing by human beings.
Brookes and coworkers~\cite{odor} proposed a quantum model to explain
odor selectivity. According to this model, molecules are recognized
not only in terms of their shape (allowing them
to dock at the right recognition site)
but also in terms of their phonon frequencies:
The molecular phonons provide the necessary energy to 
realize an inelastic electron transfer process
which takes place at the recognition site.
Such a mechanism would explain why molecules
with the same shape may have different odors
and why molecules with different shape may have the same odor.
If this is true, quantum mechanics would be fundamental
even in odor sensing. Of course, a biologist could just
assume the existence of phonons in molecules
and their coupling to charges in proteins
to kinematically explain the ``mechanics'' of odor sensing.
But the possibility of such a mechanism would come from
quantum effects anyway and this would be conceptually very important
for the understanding of life.

There are other types of phenomena in biology where
quantum coherence is also fundamental in the kinematics itself.
One such phenomenon is  the wave-like motion
of massive molecules. In a different context, the possibility
that an ensemble of massive molecules could behave as waves
and display diffraction has been experimentally proven
by Zeilinger and coworkers~\cite{zeilinger}. 
They have shown that slow beams of fullerene molecules
(${\rm C}_{60}$) can give rise to diffraction effects as much as
lighter particles (e.g., electrons and neutrons) do.
The key to understand such a phenomenon is that the de Broglie wavelength
$\lambda$ of any object is inversely proportional to its momentum $p$:
$\lambda=h/p$, 
where $h$ is the celebrated Plank's constant.
Now, the momentum is equal to the product of the mass times the velocity
of the object ($p=mv$). Therefore, even if $m$ is large, as
is the case of a fullerene molecule, a small velocity $v$ of translation of the centre
of mass can be associated with an appreciable de Broglie wavelength:
Indeed, Zeilinger has been able to measure it.
A second crucial step, necessary for the experimental measurement of $\lambda$
in fullerenes, is that decoherence~\cite{decoherence} (which is a universal and fast process,
taking place on the scales of femtosecons) must be inhibited in some way,
otherwise the wave nature of the beam would persist for 
too short times with no observable effects.
In ${\rm C}_{60}$ it turns out that the centre of mass, displaying the wave
properties, is effectively decoupled from the relative motion
of the other sixty atoms in the molecule: The sixty carbon atoms constitute
the environment of the centre of mass of the molecule and without
a coupling to the environment
(or in the presence of an important energy gap)
there is no decoherence.
Can all this be relevant to biological systems?
In biological systems one finds long and heavy carbon chains.
They typically constitute the backbone of any protein.
The motion of such long chains can be represented collectively in terms
of modes, viz., they can be expressed in Fourier frequencies describing the motion
of all the atoms in the chain globally.
The dispersion relation links the smaller frequencies
to the motion of the greater number of atoms in the chain,
so that one can say that the massive modes of the chain are slow.
As a result they can have an appreciable de Broglie wavelength.
If such slow massive modes become very weakly coupled to the environment
(i.e., the water molecules and/or other proteins around) they can exhibit
interference and diffraction effects in their motion.
This has been been theoretically proven by Tuckerman
in a study of an inter-molecular proton transfer in
Malonaldehyde. Such a study was carried out by means of a very sophisticated
first-principle simulation technique which combines
a path integral representation of the heavy carbon atoms
of the molecule and a density functional representation
of the valence electrons~\cite{tuck}.
By means of this technique Tuckerman could alternatively
represent the motion of the heavy carbon atoms
by means of quantum mechanics and classical mechanics
while always representing the transferring proton quantum mechanically.
Upon calculating the free energy barriers of the transfer process
in different cases, he was able to show the importance
of the quantum motion of the heavy atoms at $T=300$ K
for a quantitatively correct description of the phenomenon.

Another quantum effect in biological systems has been experimentally
revealed quite recently.
It has to do with the exciton propagation in photosynthentic systems
such as the Fenna-Matthews-Olson complex~\cite{photo-1}
and in photosynthetic proteins~\cite{photo-2}.
The charge propagation in such systems experimentally appears
to display a wave-like character even at relatively high temperatures,
of the order of $100 \sim 200$ K.
It has been suggested that the protein environment might, in some
way, shield the coherence of the exciton transfer process
but the detailed mechanism is not yet understood.

The last example that will be discussed here is similar
to the coherent dynamics in photosynthesis:
We refer to the coherent dynamics in chromophore molecules and
to the subsequent process of vision.
Such an example will also be used to introduce
and discuss a general characteristic of biological processes
and to propose the working hypothesis pointing to the
necessity of quantum mechanical coherence in living processes
(which has already been discussed in Section~\ref{sec:livproc}).
Chromophores are small  molecules which are tightly binded in
protein pockets. Examples are given by the p-coumaric acid,
the chromophore of the Photoactive Yellow Protein~\cite{sergi},
and by the retinal in bacteriorhodopsin~\cite{duane}.
Chromophores inside a protein can catch light.
After catching light, the energy is transformed into small atomic
rearrangements of the atoms of the chromophore:
Typically, a double bond is twisted and two hydrogen atoms
make a transition from the {\it trans} configuration
(staying on opposite sites
of the double bond) to the {\it cis} configuration
(staying on the same site of the double bond)
or vice versa, depending if one speaks of the p-coumaric
acid or retinal, respectively.
It has been experimentally demonstrated in the case of the retinal
that the chromophore dynamics
inside the protein is a coherent process~\cite{duane}.
Hence, also in this case, one finds that the protein allows, in some
way, the coherent dynamics of extended and massive systems
to take place even at room temperature.
An aspect that is not yet understood is that, while the chromophores
freely flip between the {\it trans} and {\it cis} configurations in the
vacuum, inside the protein this only happens through a photocycle
composed of various steps
(of which the {\it trans-cis} transition is only the first stage). 
The work reported in~\cite{sergi} tried to elucidate this point
in the case of the p-coumaric acid in the Photoactive Yellow Protein,
but no final conclusion could be drawn.
Proton pumping from the chromophore to the protein is also involved.
What is of interest to us is that there is currently no real explanation
for the signalling state of the whole protein after light catching.
In other words, the microscopic dynamics, photon absorption plus atomic motion
of the small chromophore molecule, must be amplified at the level
of the protein in order to cause the macroscopic signalling state (or, at
least, its first stage). Therefore,  in this
phenomenon one encounters the amplification
process that has been postulated to be the key for living processes 
in Section~\ref{sec:livproc}.
Indeed, such an amplification step could have been discussed for all
the previous examples. The question in all cases would be:
What is the physical mechanism that gives rise to the amplification process?

The hypothesis that is proposed here is that the amplification process
could be explained by a long-range, coherent polarization dynamics
of electronic degrees of freedom. These are in fact shielded
from decoherence because the interaction with the thermal 
environment may only take place by overcoming a significant energy gap:
In other words, the energy associated with thermal fluctuations at room temperature
is usually much less than the energy necessary for a transition to
the first electronic excited state.
If this were not the case, 
van der Waals forces and the hydrogen bonding, for example, 
would not be present:
The incoherent transitions of different
atoms to and from the excited state (caused by thermal fluctuations)
would destroy the coherence needed by dispersion forces.
Moreover, if the dynamics of the electron clouds is coherent,
considering that the forces on the nuclei arise from the electrons,
can one really think of the dynamics of the nuclei as incoherent?
Indeed, there are theoretical reasons that suggest that when classical degrees
of freedom interact consistently with quantum ones, they also acquire
some quantum features~\cite{kapral}.
In such a case, atomic correlations of biological functions 
in thermally disordered and crowded environments could also be interpreted
in term of quantum electron coherence.
The hypothesis that quantum coherent electron dynamics/polarization
could explain in general the amplification mechanism in biological process 
is bold and, at the moment, not substantiated by scientific evidence.
As already noted before, at this stage, it should be interpreted 
just as a working hypothesis that needs to be tested.

\section{Conclusions and perspectives}\label{sec:conclusions}

Following Kaneko's book,
a critical assessment of the usefulness of the current trends
in molecular biology has been presented. The criticism is that
the elucidation of molecules does not lead to an understanding
of life as a process. The general idea of Maturana and Varela's
autopoiesis, explaining
living systems, has been briefly sketched.
Its intrinsic circularity has been superseded by postulating
the existence of a living process, proceeding 
from the microscopic scale
to the macroscopic scale and building complexity.
Both the microscopic amplification and the increasing complexity
have been assumed to be of fundamental importance for living processes.
The amplification process, in particular, seems easier
to be modeled in a mathematical way than autopoiesis.
The concepts behind condensed matter theory and quantum mechanics
have been reviewed, emphasizing that van der Waals interactions and chemical
bonding require, in general, quantum electronic coherence 
even at room temperature.
Hence, meccano-like (electrostatically founded)
condensed matter theory has been deemed inadequate for the 
understanding of biological phenomena.
Some quantum effects in biology have also been discussed
and the suggestion that signal amplification may be explained
in terms of coherent quantum electron dynamics has been proposed:
Coherent charge distribution dynamics might be a key
to the understanding of biological matter.

Does all this mean that another paradigm of condensed matter theory
is needed in order to understand biological matter? 
Here, the affirmative answer has been defended
and it has been proposed that
long-ranged interactions and correlations must be included
from the start in the theories of biological processes.
This conclusion leads immediately to some working perspectives.
One could try to devise phenomenological computer models 
that include/postulate
long-ranged correlations in the dynamics and then use them to mimic
biological processes.
On a more fundamental level, such correlated states should arise
from first-principle theories like quantum electrodynamics.
Hence, one can embark onto the very ambitious process
of finding novel (perhaps, non-perturbative) solutions
of the ground (or first excited) states of quantum
electrodynamics in densely packed (condensed) matter systems.

One last question needs to be clearly answered here.
Should a biologist study quantum mechanics and learn all about Hilbert spaces
and linear Hermitian operators? 
This is not necessary for describing biological process in a kinematical
way (i.e., disregarding their causes) since quantum mechanics can be used in many instances as a black-box theory.
Nevertheless, one should bear in mind that 
most biological structures and signalling process
would not even exist without
quantum mechanics. This may not be important for the practice of
biology but is certainly fundamental for understanding its conceptual basis.

\section*{Acknowledgments}

This work has been funded through
a competitive research grant of the University of KwaZulu Natal.
The travel to the University of Messina (Italy) has been funded
by the National Research Foundation (NRF) of South Africa
through a Knowledge and Interchange (KIC) grant.

Useful discussions with Prof. Giacomo Tripodi and Prof. Owen de Lange
(who also carefully read the manuscript) are acknowledged.
I am particularly indebted to Prof. Paolo Giaquinta,
who did not only critically review many of the ideas
here reported (thus helping me shaping them up)
but, out of friendship,
also took upon himself the burden of a very meticolous
(and for me very precious) editing of the manuscript.


\end{document}